\documentclass[twocolumn,aps,showpacs,prb,tightenlines,amsmath,amssymb,superscriptaddress]{revtex4}
\usepackage{graphicx}
\usepackage{amssymb}
\usepackage{dcolumn}
\usepackage{amsmath}
\usepackage{bm}
\usepackage{colordvi}
\newcommand{\bgreek}[1]{\mbox{\boldmath$#1$\unboldmath}}

\begin{document}              

\title{Singlet-triplet relaxation in multivalley silicon single
  quantum dots}
\author{L. Wang}
\affiliation{Hefei National Laboratory for Physical Sciences at
Microscale,
University of Science and Technology of China, Hefei,
Anhui, 230026, China}
\affiliation{Department of Physics, 
University of Science and Technology of China, Hefei,
Anhui, 230026, China}
\author{K. Shen}
\affiliation{Hefei National Laboratory for Physical Sciences at
Microscale,
University of Science and Technology of China, Hefei,
Anhui, 230026, China}
\author{B. Y. Sun}
\affiliation{Department of Physics, 
University of Science and Technology of China, Hefei,
Anhui, 230026, China}
\author{M. W. Wu}
\thanks{Author to  whom correspondence should be addressed}
\email{mwwu@ustc.edu.cn.}
\affiliation{Hefei National Laboratory for Physical Sciences at
Microscale, University of Science and Technology of China, Hefei,
Anhui, 230026, China}
\affiliation{Department of Physics, 
University of Science and Technology of China, Hefei,
Anhui, 230026, China}

\date{\today}

\begin{abstract}

We investigate the singlet-triplet relaxation due
to the spin-orbit coupling together with the electron-phonon
scattering in two-electron multivalley silicon single quantum dots, 
using the exact
diagonalization method and the Fermi golden rule. The electron-electron Coulomb
interaction, which is crucial in the electronic
structure, is explicitly included. The multivalley effect induced by the
interface scattering is also taken into account. We first study
  the configuration with a  magnetic field in the Voigt configuration 
and identify the relaxation channel of the experimental data 
by Xiao  {\em et al.} [Phys.
Rev. Lett. {\bf 104}, 096801 (2010)]. Good agreement 
with the experiment is obtained. Moreover, we
  predict a peak in the magnetic-field dependence of the singlet-triplet 
relaxation rate induced by the anticrossing of the singlet and triplet states. 
We then work on the system with 
a magnetic field in the Faraday configuration, where the different 
values of the valley splitting  are discussed.
In the case of large
valley splitting, we find the transition rates can be effectively
manipulated by varying the external magnetic field and the dot size.
 The intriguing features  of the singlet-triplet relaxation in the vicinity
of the anticrossing point are analyzed. In the case of small
valley splitting, we find that the transition rates are much smaller
than those in the case of large valley splitting, resulting from the
different configurations of the triplet states.

\end{abstract}


\maketitle
\section{INTRODUCTION}

Spin-based qubits in semiconductor quantum dots (QDs) are
believed to be a promising candidate for scalable quantum
computation.\cite{scalable} Among different kinds of QDs, GaAs ones
have been extensively investigated in the past
decade.\cite{koppens,petta,koppens2,petta2,sasaki,meunier,cheng,shen,jiang,golovach,taylor,taylor2,hanson,hanson2,amasha,climente}
As reported, the spin decoherence, 
which is essential to know for genuine 
applications in such systems, is limited by the
hyperfine interaction\cite{paget,pikus,erlingsson,khaetskii,witzel,yao,zhang,deng,coish1,cywinski,cywinski2}
and the spin-orbit coupling (SOC)\cite{dresselhaus,rashba} together with the 
scattering.\cite{cheng,shen,jiang,climente} Recently, much
attention has been given to silicon due 
to its outstanding spin coherence
properties.\cite{culcer,li,prada,pan,liu,shaji,culcer2,xiao} Specifically,
the hyperfine coupling strength in natural silicon is two orders of 
magnitude weaker than that in
GaAs\cite{coish2} and can be further reduced by isotopic
purification.\cite{taylor3} In addition, the Dresselhaus
 SOC\cite{dresselhaus} is
absent in bulk silicon because of the existence of the bulk
inversion symmetry. Although the interfaces of a confined system can
introduce an interface inversion asymmetry
(IIA),\cite{vervoort,vervoort2,nestoklon} the
SOC due to this effect is still very small. Moreover,
the absence of the 
piezoelectric interaction makes the electron-phonon scattering much
weaker compared to that in III-V semiconductor QDs.\cite{li} 
All these features suggest a long spin decoherence time
in silicon QDs, which is of great help in realizing the operation of
logic gates and the storage of information. Furthermore, as an
indirect gap semiconductor, silicon has sixfold degenerate minima of
the conduction band. This degeneracy can be splitted by  strain or
 confinement in
quantum wells into two parts: a double-degenerate subspace of lower energy
and a fourfold-degenerate subspace of higher energy. The
presence of the interfaces can further lift the twofold degeneracy by a valley
splitting energy, which is strongly dependent on the size of the confinement
structure.\cite{boykin,friesen} Moreover, the  
correlation effects in silicon are much stronger than those in GaAs due to
the enhanced electron-electron Coulomb interaction, thanks to 
the reduced kinetic energy due to the larger effective mass. 
Thus, the physics in silicon is expected to be richer.

Very recently, spin-qubits utilizing the singlet-triplet (ST) states in
silicon QDs have been actively
investigated.\cite{culcer,li,prada,pan,liu,shaji,culcer2,xiao} Culcer 
{\em et al.}\cite{culcer} analyzed the feasibility of initialization
and manipulation of ST qubits in double QDs, concentrating on the
multivalley effect. With a large valley
splitting, the exchange coupling was explicitly investigated by Li
{\em et al.}.\cite{li} However, to the best of our knowledge, study on
the ST relaxation in silicon QDs is rather
limited.\cite{prada,pan,shaji,xiao} Prada {\em et al.}\cite{prada} calculated the
ST relaxation using the perturbation method
with the lowest few levels, which has been shown to be inadequate to
  study the ST relaxation time.\cite{shen} 
 Moreover, the Coulomb interaction was not explicitly calculated in their work.
However, the strong Coulomb interaction together with the SOC are of
 critical importance to determine the energies and
wave functions of the
 eigenstates in QDs. 
Therefore, the diagonalization approach with a large number basis functions is 
 required to guarantee the convergence of the energy spectrum and the
  ST relaxation rates.\cite{shen,climente}
In the present work, we calculate the ST relaxation in silicon single
QDs by explicitly including the Coulomb interaction
and the multivalley band structure as well as the SOC,\cite{rashba,nestoklon}
which is the key of the ST relaxation mechanism discussed in this
work. In the calculation, we employ 
the exact diagonalization 
method and calculate the ST relaxation rate
from the Fermi golden rule.\cite{cheng,shen}
We first calculate the ST relaxation rate in silicon QDs with a
  parallel magnetic field (i.e., the Voigt configuration).
Our theory successfully explains the
recent experiment, by Xiao {\em et al.}\cite{xiao} and suggests that
  the measurement corresponds the relaxation of the lowest singlet,
with the dominant channel being the one associated with the lowest
  triplet. We further predict a peak in the magnetic-field dependence
of the ST relaxation rate, resulting from the large spin mixing
 at the anticrossing point between the singlet and triplet states. 
Then we investigate the perpendicular magnetic-field (the 
Faraday configuration) dependence of the ST
 relaxation rate with different values of the valley splitting.
In  the situation of large valley splitting, the 
lowest singlet and three triplet states are all
constructed by the lowest valley state. We 
find that the transition rates can be effectively
 manipulated by the magnetic field and dot size.
We also find intriguing features in the vicinity
   of the anticrossing points.
Moreover, we compare the relative contributions of the intravalley
transverse acoustic (TA) and longitudinal acoustic (LA) 
phonons to the transition rates. In the case of
small valley splitting,  the eigenstates of the lowest two valleys
contribute. We find the ST
relaxation in this case is much slower than that in the large valley
splitting one.

This paper is organized as follows. We set up the
model and give the formalism in Sec.~II. Then in Sec.~III, we utilize the exact
diagonalization method to obtain the energy spectrum and 
calculate the ST relaxation rates. Both parallel and
  perpendicular magnetic-field dependences of the ST relaxation rates
  are studied. The behavior of the
transition rates in the vicinity of the anticrossing points is
 also discussed in
this section.  We summarize in Sec.~IV.

\section{MODEL AND FORMALISM}
In our model, we choose the lateral confinement as
$V_c(x,y)=\frac{1}{2}m_t\omega_0^2(x^2+y^2)$,
 with $m_t$ and $\omega_0$ representing the in-plane effective
 mass and the confining potential frequency.\cite{fock,darwin} The
 effective diameter can then be expressed as
 $d_0=\sqrt{\hbar\pi/m_t\omega_0}$. In the growth
 direction $[001]$, $V_z(z)$ is applied within the infinite-depth well
 potential approximation. 
 Then the single-electron Hamiltonian with  magnetic
 field ${\bf B}=B_\perp\hat{\bf z}+B_\|\hat{\bf x}$
 is described by 
\begin{equation}
 H_{\rm e}=\frac{{P_x}^2+{P_y}^2}{2m_t}+\frac{{P_z}^2}{2m_z}+V({\bf
   r})+H_{\rm so}({\bf P})+H_{\rm Z}+H_{\rm v},
\label{eq1}
 \end{equation}
with $m_z$ representing the effective mass along the $z$-direction.
 $V({\bf r})=V_c+V_z$ and ${\bf P}=-i\hbar{\bgreek \nabla}+(e/c){\bf A}$,
 with ${\bf A}=(-yB_\perp,xB_\perp,2yB_\|)/2$. 
$H_{\rm so}$ describes the SOC Hamiltonian, 
 including the Rashba term\cite{rashba} due to the 
 structure inversion asymmetry and the term due to the
 IIA.\cite{vervoort,vervoort2,nestoklon}  Then, one obtains 
 \begin{equation}
 H_{\rm
   so}=a_0(P_x\sigma_y-P_y\sigma_x)+b_0(-P_x\sigma_x+P_y\sigma_y),
 \label{eq2}
 \end{equation}
 where $a_0$ and $b_0$ stand for the strengths of the Rashba and 
 IIA terms, respectively.
 The Zeeman splitting is given by
 $H_{\rm Z}=\frac{1}{2}g\mu_B(B_\perp\sigma_z+B_\|\sigma_x)$ 
 with $g$ being the Land\'{e} factor. Since the four in-plane valleys are
 separated from the two out-of-plane ones by a large energy gain, only
 the two out-of-plane valleys are relevant  in the calculation. $H_{\rm v}$ in
 Eq.\,(\ref{eq1}) describes the coupling\cite{boykin,friesen} between
 the valleys lying at $\pm\langle k_0\rangle$ 
 along the $z$-axis with $\langle k_0\rangle=0.85(2\pi/a_{\rm
   Si})$. Here, $a_{\rm Si}=5.43\,\mathring{\rm A}$ is the
 lattice constant of silicon. Throughout this paper, one uses the
 subscripts ``$z$'' and ``$\bar{z}$'' to denote the valley at
 $\langle k_0\rangle$ and the one at $-\langle k_0\rangle$, respectively.

 By solving the Schr\"odinger equation of the Hamiltonian
 $H_0=\tfrac{1}{2m_t}({P_x}^2+{P_y}^2) +\tfrac{1}{2m_z}{{P_z}^2}+V({\bf 
   r})$, one obtains the eigenvalues\cite{fock,darwin}
 \begin{equation}
 E_{nln_{z}}=\hbar\Omega(2n+|l|+1)+\hbar
 l\omega_B+\frac{{n_z}^2{\pi}^2{\hbar}^2}{8m_za^2},
 \label{eq3}
 \end{equation}
 where $\Omega=\sqrt{{\omega_0}^2+{\omega_B}^2}$ and
  $\omega_B={eB_\perp}/{(2m_t)}$.  $a$ represents the half-well 
   width. Here, $n=0,1,2,...$ is the radial 
 quantum number and $l=0,\pm 1,\pm 2,...$ represents the azimuthal angular
 momentum quantum number. The index $n_z$ denotes the subbands
 resulting from the confinement along the growth direction.
 The corresponding eigenfunctions read
 \begin{eqnarray}
   \nonumber
   F_{nln_{z}}({\bf r})&=&N_{n,l}(\alpha r)^{|l|}e^{-(\alpha
     r)^{2}/2}e^{il\theta}L_{n}^{|l|}{\big (}(\alpha r)^2{\big )}\\
   && \mbox{}\times  \left\{
     \begin{array}{ll}
      \frac{1}{\sqrt{a}}\sin[\frac{n_z\pi}{2a}(z+a)], &\mbox{$|z|\le
        a$}\\
      0, &\mbox{otherwise},
    \end{array}
    \right.
    \label{eq4}
 \end{eqnarray}
 with $N_{n,l}=\{{\alpha}^2n!/[\pi(n+|l|)!]\}^{1/2}$ and
 $\alpha=\sqrt{m_t\Omega/\hbar}$. $L_{n}^{|l|}$ is the generalized
 Laguerre polynomial. The wave functions in different valleys can
then be expressed as 
 $\phi_{nln_z}^{z,\bar{z}}=F_{nln_{z}}({\bf r})e^{\pm ik_{0}z}u_{z,\bar{z}}({\bf
   r})$, with $u_{z,\bar{z}}({\bf r})$ representing the lattice-periodic Bloch
 functions.\cite{culcer} Here, we neglect the orbital effect of
   the parallel magnetic field by considering the strong confinement
   along the growth direction.
 
 One can demonstrate that the overlap between the wave functions in
 different valleys is negligibly small, therefore only $H_{\rm v}$ is
 considered to contribute to the intervalley coupling in the present
 work. However, there still remain some
   controversies over the valley coupling
   nowadays.\cite{friesen,saraiva,chutia} In this work, we
take $\langle\phi_{nln_z}^{z,\bar{z}}|H_{\rm
   v}|\phi_{n^\prime l^\prime
   n_z}^{\bar{z},z}\rangle=\Delta^1_{n_z}\delta_{nn^\prime}\delta_{ll^\prime}$
 and   $\langle\phi_{nln_z}^{z,\bar{z}}|H_{\rm
   v}|\phi_{n^\prime l^\prime
   n_z}^{z,\bar{z}}\rangle=\Delta^0_{n_z}\delta_{nn^\prime}\delta_{ll^\prime}$,
   according to Ref.\,\onlinecite{friesen}. Here,  only the  
 coupling element between the states with identical $n_z$ is
 given, since only the first subband is included in our calculation
 while the others are neglected due to the much higher energy. Including
 this intervalley coupling, the eigenstates become
 $\phi_{nln_z}^{\pm}=\frac{1}{\sqrt{2}}(\phi_{nln_z}^z\pm \phi_{nln_z}^{\bar{z}})$ with
 eigenvalues $E_{nln_z}^{\pm}=E_{nln_{z}}+\Delta_{n_z}^0\pm |\Delta_{n_z}^1|$. 
In these equations, 
 \begin{eqnarray}
 \Delta_{n_z}^0&=&\frac{V_{\rm v}{n_z}^2{\pi}^2{\hbar}^2}{4m_za^3},\\
 \Delta_{n_z}^1&=&\frac{V_{\rm
     v}{n_z}^2{\pi}^2{\hbar}^2\cos(2k_0a)}{4m_za^3},
 \label{eq6}
 \end{eqnarray} 
with $V_{\rm v}$ standing for the ratio
 of the valley coupling strength to the depth of quantum well.\cite{friesen}

 For a two-electron QD, the total Hamiltonian is given by 
 \begin{equation}
 H_{\rm tot}=(H_{\rm e}^1+H_{\rm e}^2+H_{\rm C})+H_{\rm p}+H_{\rm ep}^1+H_{\rm
   ep}^2.
 \label{eq7}
 \end{equation}
Here, the two electrons are labeled by $``1"$ and $``2"$.  The
electron-electron Coulomb interaction is described by
 $H_{\rm C}=\frac{e^2}{4\pi\epsilon_0\kappa|{\bf r_1}-{\bf
     r_2}|}$ with $\kappa$ representing the relative static dielectric
 constant. $H_{\rm p}=\sum_{{\bf q}\lambda}\hbar\omega_{{\bf
     q}\lambda}a_{{\bf q}\lambda}^+a_{{\bf q}\lambda}$ represents the
 phonon Hamiltonian with $\lambda$ and ${\bf q}$ denoting the phonon
 mode and the momentum respectively. The electron-phonon interaction
 Hamiltonian is given by $H_{\rm ep}=\sum_{{\bf q}\lambda}M_{{\bf
     q}\lambda}(a_{{\bf q}\lambda}^++a_{-{\bf q}\lambda})e^{i{\bf
     q}\cdot{\bf r}}$.

We construct two-electron basis functions in the forms of
 either singlet or triplet based on the the single-electron
 eigenstates. For example, we use two single-electron spatial wave 
 functions $|n_1l_1n_{z1}n_{v1}\rangle$ and $|n_2l_2n_{z2}n_{v2}\rangle$ (denoted as
 $|N_1\rangle$ and $|N_2\rangle$ for short; $n_v=\pm$) to obtain the singlet
 functions
\begin{equation}
  |S^{(\Xi)}\rangle=(|\uparrow\downarrow\rangle
-|\downarrow\uparrow\rangle)\otimes
  \begin{cases}
    \frac{1}{\sqrt 2}|N_1N_2\rangle,& N_1=N_2\\
    \frac{1}{2}(|N_1N_2\rangle+|N_2N_1\rangle)
    ,& N_1\not=N_2,
  \end{cases}
  \label{eq8}
\end{equation} 
and the triplet functions for $N_1\ne N_2$
\begin{eqnarray}
 &&|T_+^{(\Xi)}\rangle=\frac{1}{\sqrt2}(|N_1N_2\rangle-|N_2N_1\rangle)\otimes|\uparrow\uparrow\rangle,\\
 &&|T_0^{(\Xi)}\rangle=\frac{1}{2}(|N_1N_2\rangle-|N_2N_1\rangle)\otimes(|\uparrow\downarrow\rangle
 +|\downarrow\uparrow\rangle),\\
 &&|T_-^{(\Xi)}\rangle=\frac{1}{\sqrt2}(|N_1N_2\rangle-|N_2N_1\rangle)\otimes|\downarrow\downarrow\rangle.
 \label{eq9_11}
\end{eqnarray} 
Here, the spatial wave functions of the first and
second electrons in $|NN'\rangle$ are denoted as $N$ and $N'$ in
sequence. The superscript $(\Xi)$ denotes the valley
  configuration of each state. We define $\Xi=\pm$ for the valley indexes
  of single electron states $n_{v1}=n_{v2}=\pm$, and $\Xi=m$ for 
$n_{v1}\ne n_{v2}$.

Then, one can calculate the matrix elements of the Coulomb
interaction, which can be expressed by
\begin{widetext}
 \begin{equation}
\langle N_1N_2|H_{\rm C}|N_1^{\prime}N_2^{\prime}\rangle=\frac{e^2}{16{\pi}^2\epsilon_0\kappa}
\delta_{l_1+l_2,l_1^{\prime}+l_2^{\prime}}
\sum_{\gamma_{1},\gamma_{2},\gamma_1^{\prime},\gamma_2^{\prime}=z,\bar{z}}
\eta_{n_{v1}}^{\gamma_{1}}\eta_{n_{v2}}^{\gamma_{2}}
\eta_{n_{v1}^\prime}^{\gamma_1^{\prime}}
\eta_{n_{v2}^\prime}^{\gamma_2^{\prime}}
G(\phi_{n_1l_1n_{z1}}^{\gamma_1},\phi_{n_2l_2n_{z2}}^{\gamma_2},
\phi_{n_1^\prime l_1^\prime n_{z1}^\prime}^{\gamma_1^\prime},
\phi_{n_2^\prime l_2^\prime n_{z2}^\prime}^{\gamma_2^\prime}),
 \end{equation}
 \label{eq12}
\end{widetext}
where the superscripts $\gamma_i$ and $\gamma_i^\prime$ run over the two valleys,
$z$ and $\bar z$, with $\eta_\pm^z=1$ and
$\eta_+^{\bar z}=-\eta_-^{\bar z}=1$.
$G$ is given in detail in Appendix. One also calculates
the SOC and Zeeman splitting terms,
hence obtains the two-electron Hamiltonian, i.e., the terms
in the bracket in Eq.\,(\ref{eq7}). Then, the two-electron
eigenvalues and eigenfunctions can be obtained by exactly
diagonalizing the two-electron Hamiltonian. We identify a two-electron
eigenstate as  singlet (triplet) if its amplitude of the singlet (triplet) 
components is larger than 50~\%.

From the Fermi golden rule, one can
calculate the transition rate from the state
$|i\rangle$ to $|f\rangle$, due to the 
electron-phonon scattering,
\begin{eqnarray}
  \Gamma_{i\rightarrow f}&=&\frac{2\pi}{\hbar}\sum_{{\bf
      q}\lambda}|M_{{\bf q}\lambda}|^2|\langle f|\chi|i\rangle|^2[\bar{n}_{{\bf
      q}\lambda}\delta(\epsilon_f-\epsilon_i-\hbar\omega_{{\bf
      q}\lambda})\nonumber\\
  &&\mbox{}+(\bar{n}_{{\bf
      q}\lambda}+1)\delta(\epsilon_f-\epsilon_i+\hbar\omega_{{\bf
      q}\lambda})],
  \label{eq13}
\end{eqnarray}
in which $\chi({\bf q},{\bf r_1},{\bf r_2})=e^{i{\bf q}\cdot{\bf
    r_1}}+e^{i{\bf q}\cdot{\bf r_2}}$ and
$\bar{n}_{{\bf q}\lambda}$ stands for the Bose distribution of phonons. In
our calculation, the temperature is fixed at 0\ K. Thus $\bar{n}_{{\bf
    q}\lambda}=0$ and only the second term contributes.


\section{NUMERICAL RESULTS}
Since the piezoelectric interaction is absent in silicon\cite{li} and
the energy difference between the initial and final 
states discussed here is much smaller than the energies of the
intervalley acoustic phonon and the
optical phonon,\cite{pop} one only needs to calculate the intravalley
electron-acoustic phonon scattering due to the deformation
potential. In the present work, both the TA 
and LA phonons are included. The corresponding matrix
elements are $M_{\beta,{\rm intra},{\bf
    Q}}^2={\hbar D_{\beta}^2Q^2}/{(2d\Omega_{\beta,{\rm intra},{\bf
      Q}})}$ with $\beta$=LA/TA standing for the LA/TA phonon
mode. Here, we take the mass density of silicon $d=2.33$\,g/cm$^3$.\cite{sonder}
The deformation potentials for the LA and TA phonons are $D_{{\rm LA}}=6.39$\,eV
and $D_{{\rm TA}}=3.01$\,eV, respectively.\cite{pop} 
The phonon energy $\Omega_{\beta,{\rm intra},{\bf Q}}=v_{\beta}Q$ with
sound velocities $v_{\rm LA}=9.01\times 10^5$\,cm/s and $v_{\rm
  TA}=5.23\times 10^5$\,cm/s.\cite{pop} 
In our calculation, we take $m_t=0.19m_0$ and $m_z=0.98m_0$ with $m_0$
being the free electron mass.\cite{dexter} The Land\'{e} factor
$g=2$,\cite{graeff} the ratio 
$V_{\rm v}=7.2\times 10^{-11}$\,m.\cite{friesen} In the calculation,
  we employ the exact diagonalization 
method with the lowest $1516$ singlet and $4452$ triplet basis functions
to guarantee the convergence of the energies and the transition
rates. 

One finds that the eigenstates composed by the two-electron
  basis functions with single valley state  ``$-$'' are
almost independent of those constructed by the 
ones with single valley state ``$+$'' and two valley
  states ``$-$'' and ``$+$''. On the one
hand, there is nearly no coupling between them due to the
negligibly small intervalley Coulomb
interaction\cite{li} and overlap between the
wave functions in different
valleys. One can also demonstrate that the
elements of the SOCs between the
states with different valley indices vanish when only the first
subband is included, regardless of the coupling
strengths $a_0$ and $b_0$. On the other hand, the transition between
them is almost forbidden because
$\langle f|\chi|i\rangle$ in Eq.\,({\ref{eq13}}) is strongly
suppressed thanks to the large intervalley wave vector $\langle
2k_0\rangle$ from the difference of the phases between different
valleys.  Therefore, the eigenstates are divided into three
independent sets based on the valley indexes. It is noted that the energy of
the eigenstate with valley configuration ``$-$'' is smaller than the
corresponding levels with valley configurations ``$+$'' and ``$m$'' due to
  the contribution of the valley splitting.

\begin{figure}[bth]
\includegraphics[width=7.5cm]{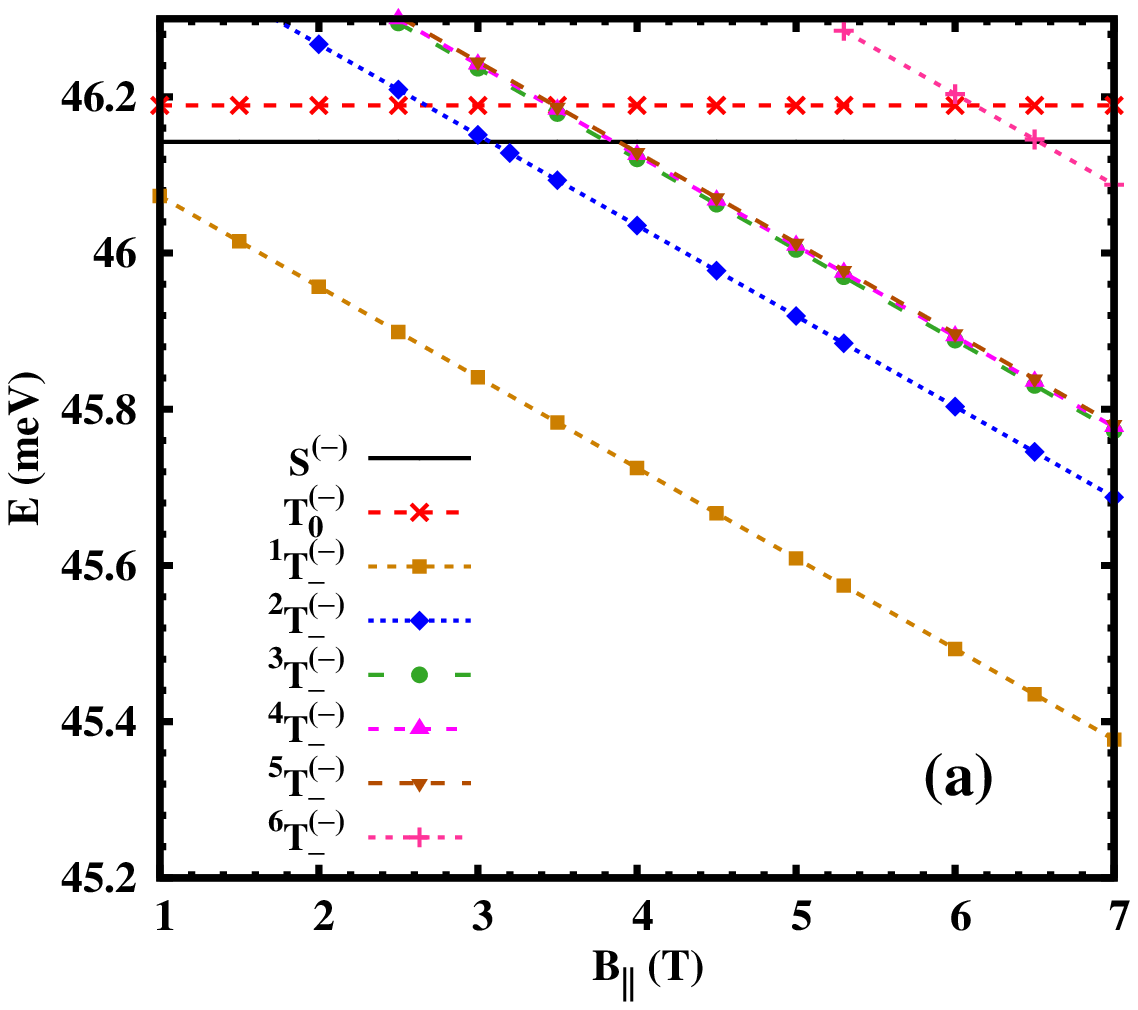}
\includegraphics[width=7.5cm]{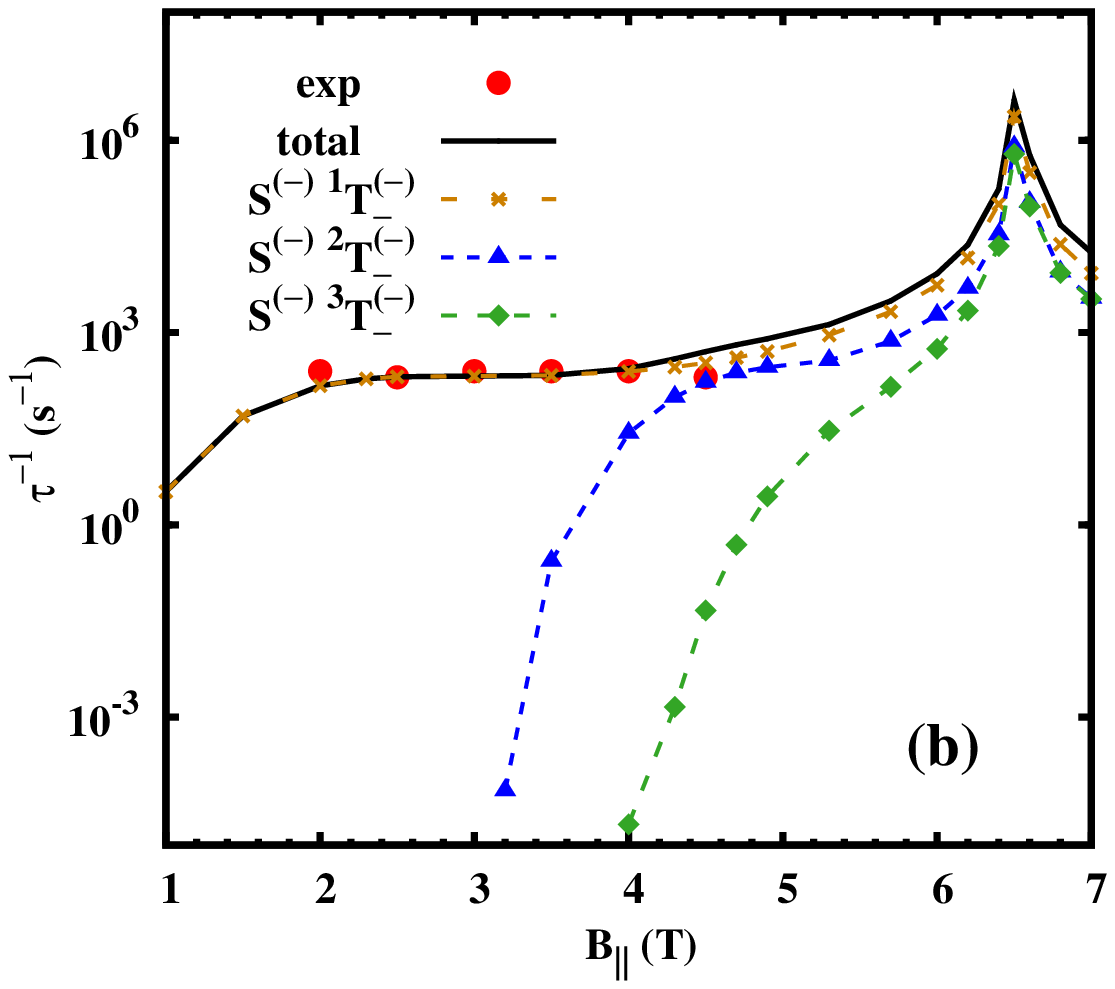}
\caption{(Color online) (a) The lowest few energy levels {\em vs}.
  parallel magnetic field $B_{\|}$ in a single QD. Note that each of
$|T_0^{(-)}\rangle$ and $|^{i}T_-^{(-)}\rangle$ ($i=1,2,3,6$)  is
 double-degenerate. (b) Relaxation rates {\em vs}. the   
magnetic field. The red dots stand for the
  experimental
data. ${\rm S^{(-)}\ }$$^i$${\rm T_-^{(-)}}$ ($i=1$-$3$) denotes the sum of the
relaxation rates from $|S^{(-)}\rangle$ to the two
degenerate $^i$${\rm T_-^{(-)}}$ levels. 
In the calculation, $2a=4.344$\,nm and $d_0=56$\,nm.}  
\label{fig1}
\end{figure}

\subsection{PARALLEL MAGNETIC-FIELD DEPENDENCE}
Very recently, Xiao {\em et al.}\cite{xiao}
measured the ST relaxation time in Si/SiO$_2$ QDs under a
magnetic field parallel to the interface of the heterostructure.
They reported that the ST relaxation time only 
slightly fluctuates around 5~ms when the magnetic field increases
from 2 to 4.5~T. In the experiment, the orbital level
spacing is observed to be $\sim0.4$\,meV, corresponding to the
 effective diameter of the QD $d_0=56$\,nm. 
However, some parameters such as the effective well width,
the strengths of SOCs and the valley
 splitting are unavailable. Moreover, 
the channel of the relaxation process is not identified because of
the uncertainty of the exact excited states spectrum in the
experiment.\cite{xiao} Here, we take advantage of our model to
clarify the experiment situation. In the
calculation, we assume the magnetic field 
  along $x$-direction and take the relative static dielectric constant
  $\kappa=7.9$.\cite{li} 
Since the valley splitting is strongly
  dependent on the effective well width according to Eq.\,(6),
  it is difficult to determine the energy spectrum without the
  knowledge of the exact well width. For a large valley splitting,
  the lowest levels are all constructed by the states with the single
  valley index ``$-$'', and the energy difference between the 
adjacent levels is determined solely by the
  orbital level spacing and Zeeman splitting 
approximately. Therefore, the relaxation rate of each
 excited state can be calculated to identify the
  relaxation channel in the experiment. 
However, the lowest levels become more complicated for a small valley
splitting because new levels with the valley index ``$+$'' become relevant.
Fortunately, as said above, 
the inclusion of the states with valley configuration ``$+$'' or
``$m$'' has no observable influence on the relaxation 
of the states with the
valley configuration ``$-$''. In the following, 
we first study the large valley  splitting case.
We take 32 monoatomic layers of silicon
    along the growth direction, corresponding to the
well width $2a=4.344$\,nm ($2|\Delta_{n_z}^1|=0.83$\,meV). 
The strengths of the Rashba SOC and
IIA term are used as fitting parameters. We first calculate the energy
spectrum since it is weakly dependent on the strengths  of the SOCs. The
 lowest few levels, denoted as $|S^{(-)}\rangle$,
$|T_0^{(-)}\rangle$ and $|^iT_-^{(-)}\rangle$ (spin down) ($i=1$-$6$) according 
to their major components, are plotted as function of the magnetic field in
Fig.\,\ref{fig1}(a).  As the magnetic field increases, the energies of
$|S^{(-)}\rangle$ and $|T_0^{(-)}\rangle$  
keep invariant while those of $|^iT_-^{(-)}\rangle$ ($i=1$-$6$)
decrease due to the Zeeman splitting. 
The major component of $|S^{(-)}\rangle$ is 
constructed by the single-electron states
$|01\rangle|0-1\rangle$ according to Eq.\,(\ref{eq8}), and
those of the triplet states $|^4T_-^{(-)}\rangle$ and $|^5T_-^{(-)}\rangle$
are given by $|01\rangle|0-1\rangle$ and $|00\rangle|10\rangle$
following Eq.\,(\ref{eq9_11}), respectively. We find $|^iT_-^{(-)}\rangle$
($i=1,2,3,6$) and $|T_0^{(-)}\rangle$ are all double-degenerate.
 The major components of the two levels of $|^1T_-^{(-)}\rangle$,
 $|^2T_-^{(-)}\rangle$, $|^3T_-^{(-)}\rangle$, and $|^6T_-^{(-)}\rangle$ are
 composed by the single-electron states $|00\rangle|0\pm 1\rangle$,
 $|0\pm 1\rangle|0\pm 2\rangle$, $|00\rangle|0\pm 2\rangle$ and 
 $|10\rangle|1\pm 1\rangle$ in sequence. $|T_0^{(-)}\rangle$ is mainly
 constructed by the basis function involving $|00\rangle|0\pm 1\rangle$ also.
Here, we only retain the quantum numbers
$n$ and $l$ for short, because other quantum numbers of these single-electron
states are all the same.
 
We then calculate the relaxation rates of these states due to
 phonon emission. Due to the low temperature in the
  experiment,\cite{xiao} the relaxation rate at zero temperature
can well represent the experimental data. We find that if one takes the
Rashba SOC strength $a_0=-2.09$\,m/s and the IIA term strength
$b_0=-10.44$\,m/s, the total relaxation rate of the state
$|S^{(-)}\rangle$ fits the experimental  data pretty well as shown in
Fig.\,\ref{fig1}(b) (from 2 to 4.5 Tesla). The relaxation
rates of other levels can not recover the experiment
results. Specifically, the relaxation rate of
  $|T_0^{(-)}\rangle$ presents a peak at $B_{\|}\sim 3.45$\,T (not shown in
  the figure) and those of $|^iT_-^{(-)}\rangle$ ($i=2$-$6$) 
relax too fast (in the
  magnitude of $\sim 1$\,ns). Therefore, we conclude that the experimental
data might correspond to the lifetime of the singlet $|S^{(-)}\rangle$. 
The rates of the major relaxation channels of  
$|S^{(-)}\rangle$ (involving ${\rm S^{(-)}}$ $^i$${\rm
  T_-^{(-)}}$, $i=1$-$3$) are also plotted in
Fig.\,\ref{fig1}(b). Interestingly, the calculation predicts a peak of 
the total relaxation
rate at $B_{\|}\sim 6.5$\,T, which should be checked by
future experiments. Moreover, one also finds the significant increase of the total
relaxation rate by increasing the magnetic field in
the small magnetic field regime, i.e., below 2~T. Such  rich
magnetic-field dependences can be understood as follows. 
From the figure, we find that the dominant relaxation channel is the
  one from $|S^{(-)}\rangle$ to $|^1T_-^{(-)}\rangle$.
In the small magnetic field regime, the energy of the phonon emmision
of this channel (corresponding to the energy difference between
$|S^{(-)}\rangle$ and $|^1T_-^{(-)}\rangle$) is small and linearly increases with the magnetic field, which lead to the significant enhancement of the
transition.\cite{climente,shen} However, the transition rate becomes
insensitive to the phonon energy since the value of
$\langle f|\chi|i\rangle$ in Eq.\,({\ref{eq13}}) is suppressed for a
large phonon momentum, then the transition rate only slightly varies beyond
 2~T. This picture can be also used to understand the feature of the 
relaxation rates between
$|S^{(-)}\rangle$ and $|^iT_-^{(-)}\rangle$ ($i=2,3$) far away from the peak.
The peak at $B_{\|}\sim 6.5$\,T, where the triplet state
$|^6T_-^{(-)}\rangle$ intersects the singlet state $|S^{(-)}\rangle$, results from the
strong coupling between them due to the SOCs. To ease further
discussion, one denotes the total angular momentum
and spin states as $L=l_1+l_2$ and $(S,S_x)$, respectively, with $S_x$
representing the $x$ component of the total 
spin ${\bf S}$. By neglecting the terms with $\sigma_x$ in
Eq.\,(\ref{eq2}) due to its
smaller magnitude compared with the Zeeman splitting, one obtains the
SOC Hamiltonian
\begin{equation}
  H_{\rm so}=\left[{a_0}(P^++P^-)-ib_0(P^+-P^-)\right](S^++S^-)/\hbar,
  \label{eq14}
\end{equation}
with $S^{\pm}=S_y\pm iS_z$.
The ladder operations $P^{\pm}$ and $S^{\pm}$ change $L$ and
$S_x$ by one unit, respectively. Therefore, a state with $(L,S_x)$ can
couple with the one with $(L\pm 1,S_x\pm 
1)$ or $(L\pm 1,S_x\mp 1)$ for both the Rashba and IIA terms. 
From the major components of the two-electron eigenstates, the quantum
numbers $(L,S_x)$ of $|^6T_-^{(-)}\rangle$ 
and $|S^{(-)}\rangle$ are $(\pm 1,-1)$ and $(0,0)$, respectively. It
is obvious that $|^6T_-^{(-)}\rangle$ directly
couples with $|S^{(-)}\rangle$ through the SOCs. 
As a result, there is an energy gap (too
tiny to pick up in the figure) at the intersecting point between
$|^6T_-^{(-)}\rangle$ and $|S^{(-)}\rangle$, which means an anticrossing event
occurs. In the vicinity of this anticrossing point, the wave  function
of $|S^{(-)}\rangle$ contains a large amount of the spin-down triplet
component, which enhances the spin relaxation process. One notices that the
intersecting point between $|^2T_-^{(-)}\rangle$ and $|S^{(-)}\rangle$ is also an
anticrossing point. However, the coupling between these states is
indirect and small, hence only slightly affects the ST
relaxation. Other intersecting points between $|S^{(-)}\rangle$ and
$|^iT_-^{(-)}\rangle$ ($i=3$-$5$) are just simply crossing points.
Moreover, one finds that the peak of the
relaxation rate of $|T_0^{(-)}\rangle$ at $B_{\|}\sim 3.45$\,T reflects the
anticrossing behavior between $|T_0^{(-)}\rangle$ and $|^iT_-^{(-)}\rangle$
  ($i=3$-$5$) and the fast relaxation of $|^iT_-^{(-)}\rangle$ ($i=2$-$6$)
  comes from the same spin configuration of the initial and final
  states. As discussed above, the relaxation rates
  of the states with single valley state ``$-$'' are 
  insensitive to the valley splitting. Therefore, the results  in the
case of small valley splitting  are almost the same as the case of 
large valley splitting. Moreover, we find the 
  results are also robust against the effective well width.

Similarly, one can calculate the relaxation rates of the another set of
 states with the valley configuration ``$+$''. The total relaxation
rate of $|S^{(+)}\rangle$ can {\em also} recover the experimental data
pretty well, where the
channel between $|S^{(+)}\rangle$ and $|^1T_-^{(+)}\rangle$ is the
dominant one. Here, $|S^{(+)}\rangle$ and $|^1T_-^{(+)}\rangle$ are
the lowest singlet and triplet states of this set of
valley configuration, separately. 
As for the set composed by the states with the valley
configuration ``$m$'', more triplet basis functions (constructed by two
single-electron basis functions with the same quantum numbers $n$ and
$l$) should be included. This makes the results of this set of states 
different from
the other two with single valley index ``$-$'' or ``$+$''.
However, as the energies of the states with single valley state ``$+$'' 
are higher than the corresponding ones with valley state ``$-$'', we
suppose the experimental data by Xiao {\em et al.}\cite{xiao} corresponding to
the states with ``$-$'' valley index.

\begin{figure}[bth]
\includegraphics[width=7.5cm]{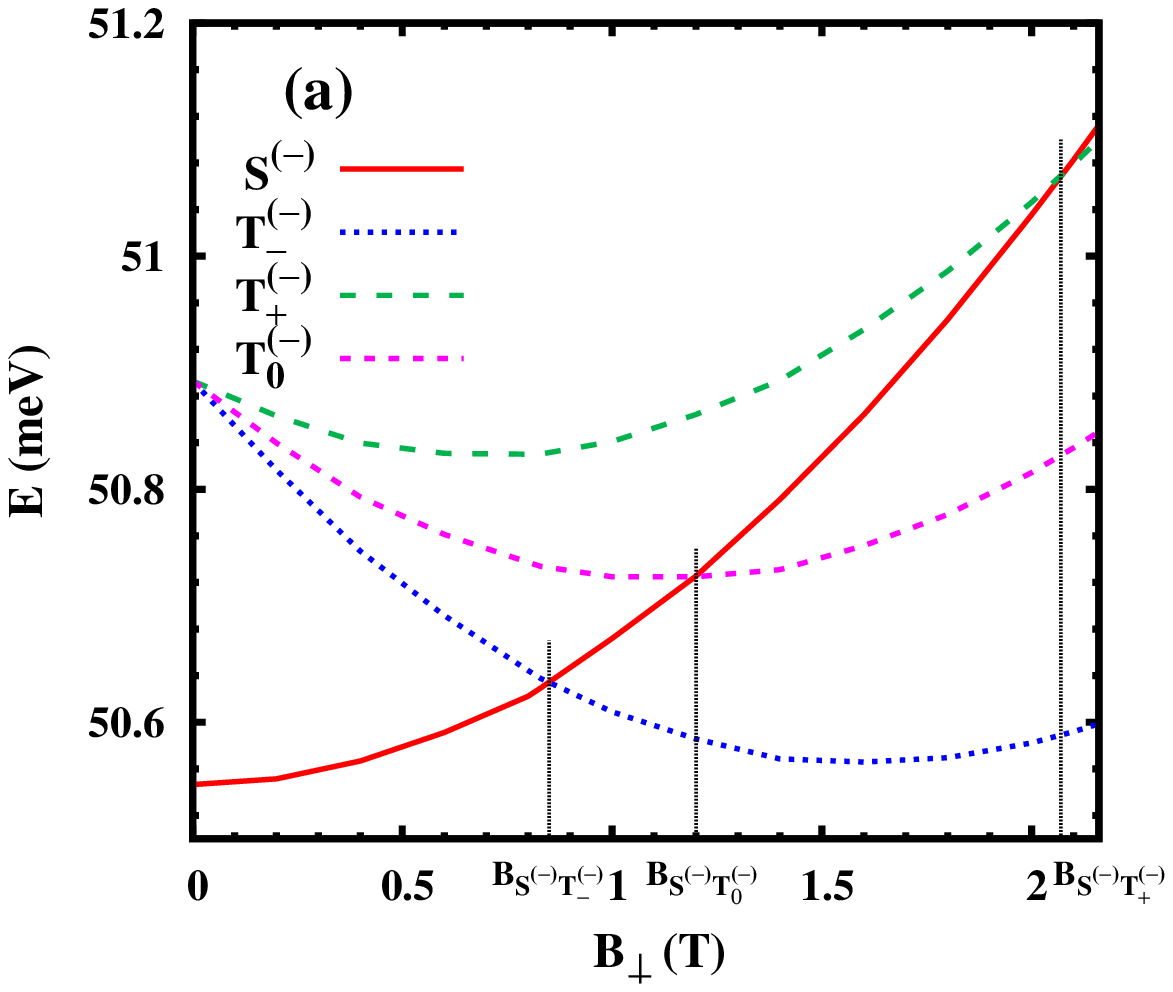}
\includegraphics[width=7.5cm]{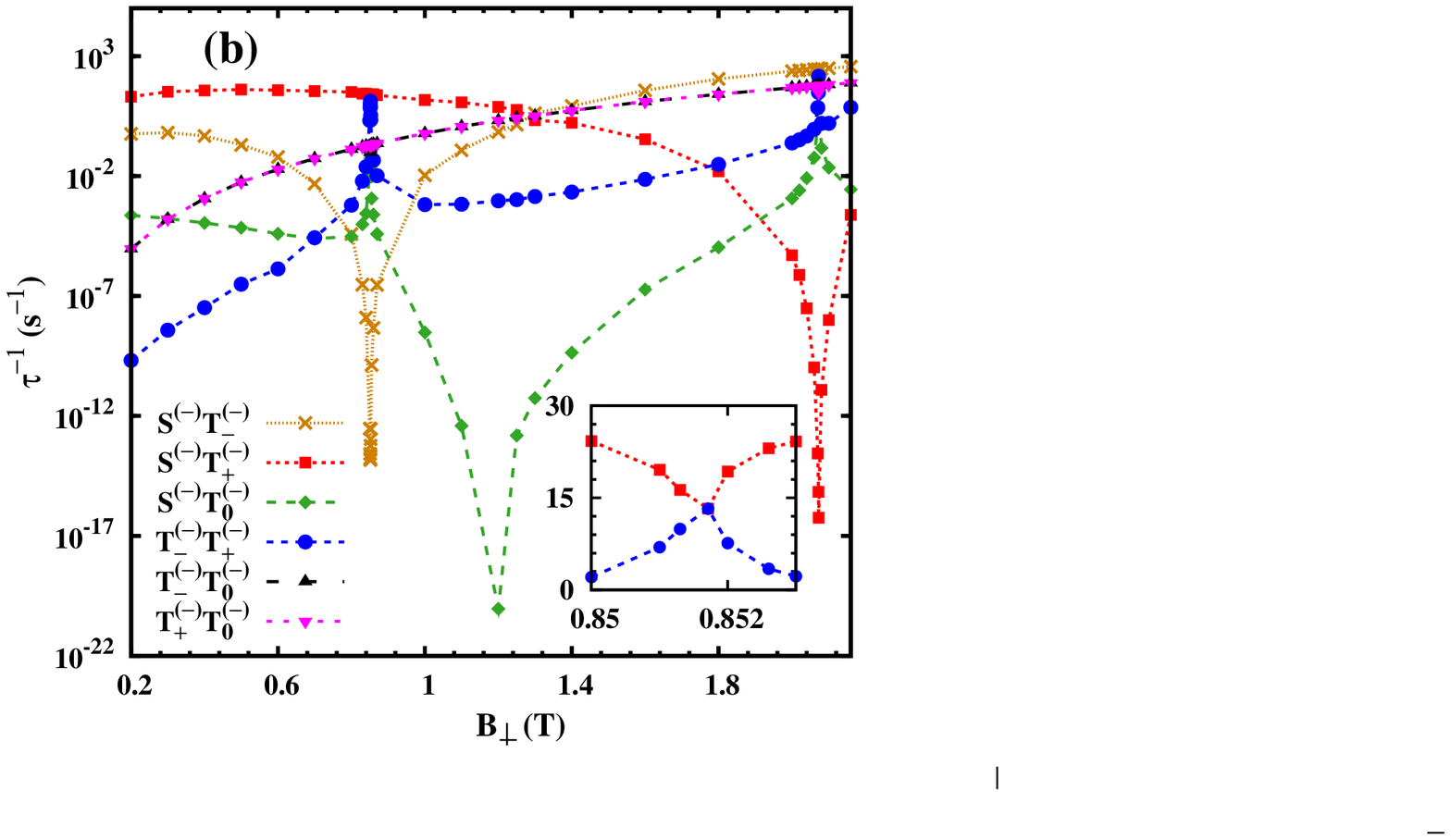}
\caption{(Color online) (a) The lowest four energy levels {\em vs}.
  perpendicular magnetic field $B_{\perp}$ in a 
single QD. The anticrossing/crossing points
  are labeled as $B_{\rm S^{(-)}T_-^{(-)}}$, $B_{\rm S^{(-)}T_0^{(-)}}$ and $B_{\rm S^{(-)}T_+^{(-)}}$. 
(b) Transition rates {\em vs}. the   
magnetic field. The inset zooms the range near $B_{\rm S^{(-)}T_-^{(-)}}$ with
  the rates of channels ${\rm S^{(-)} T_+^{(-)}}$ and ${\rm T_-^{(-)} T_+^{(-)}}$. In the
  calculation, $2a=4.344$\,nm and $d_0=29$\,nm.}  
\label{fig2}
\end{figure}

\subsection{PERPENDICULAR MAGNETIC FIELD DEPENDENCE}
In this part, we turn to the perpendicular magnetic field case and choose
 SiGe/Si/SiGe QDs without loss of
  generality. The relative static dielectric constant is
  $\kappa=11.9$ in this structure.\cite{sze} We start from the structure
  with 32 monoatomic layers of silicon along the growth direction of the 
quantum well as in the parallel magnetic field case,
corresponding to a large valley splitting $2|\Delta_{n_z}^1|=0.83$\,meV.
With an electric field 30\,kV/cm along the growth direction, 
one obtains the strength of the Rashba SOC
induced by this electric field $a_0=\mp6.06$\,m/s and that of the IIA term
$b_0=\mp30.31$\,m/s for the SOC elements between the states with identical 
valley index ``$\mp$''.\cite{nestoklon}
Moreover, a large effective diameter $d_0=29$\,nm is taken to
ensure that the lowest levels are constructed only by the basis
 functions with valley index ``$-$''.

 The first four levels in the QD are plotted in Fig.~\ref{fig2}(a) 
as function of the perpendicular magnetic field.
They are labeled as $|S^{(-)}\rangle$,
$|T_+^{(-)}\rangle$ (spin up), $|T_0^{(-)}\rangle$
and $|T_-^{(-)}\rangle$ (spin down),
according to their major components.  The shape of the spectrum
  can be understood from the single-electron spectrum of
  Eq.\,(\ref{eq3}). For example, the major component of
$|S^{(-)}\rangle$, i.e., $|S^{1(-)}\rangle$, is composed by two electrons
in $|001-\rangle$ state, hence the magnetic-field dependence of
$\epsilon_{S^{(-)}}$
is given by $2E_{001}^-$ approximately. Similarly, the magnetic-field
dependence of the triplet $|T_0^{(-)}\rangle$ ($|T_\pm^{(-)}\rangle$) can be
described by $E_{001}^-+E_{0 
-11}^-$ ($E_{001}^-+E_{0-11}^-\pm E_Z$, with $E_z$ representing the
Zeeman splitting), because this state mainly contains the triplet basis
$|T^{1(-)}_0\rangle$ which involves the single-electron functions $|001-\rangle$ and
$|0{-1}1-\rangle$. The qualitative analysis still works even with
the strong Coulomb interaction. It is shown that the singlet state $|S^{(-)}\rangle$
intersects the three triplet levels with the increase of the magnetic
field. Since the crossing and/or anticrossing points show different properties
on ST relaxation as discussed above, we now analyze the
intersecting points. We still denote the two-electron angular momentum
as $L=l_1+l_2$, but take the spin states $(S,S_z)$ instead by
considering the perpendicular magnetic field.
The SOC Hamiltonian can be rewritten as\cite{climente} 
\begin{equation}
  H_{\rm so}=\frac{2ia_0}{\hbar}(P^+S^--P^-S^+)-\frac{2b_0}{\hbar}(P^+S^++P^-S^-),
  \label{eq15}
\end{equation}
with the ladder operations $P^{\pm}$ and $S^{\pm}$ changing $L$ and $S_z$ by one unit,
respectively. Here, $S^{\pm}=S_x\pm iS_y$.
It is clear that a state with $(L,S_z)$ can couple
with the one with $(L\pm 1,S_z\mp 1)$ due to the Rashba SOC and 
the one with $(L\pm 1,S_z\pm 1)$ due to the IIA term.
Approximately, the quantum numbers $(L,S_z)$ of $|T_-^{(-)}\rangle$ and
  $|S^{(-)}\rangle$ are $(-1, -1)$ and $(0,0)$, respectively, according to
  the wave functions of $|T_-^{1(-)}\rangle$ and $|S^{1(-)}\rangle$.
Therefore, the IIA term couples these states
and an anticrossing event occurs at the intersecting point
 ($B_{\rm S^{(-)}T_-^{(-)}}\sim 0.85$\,T), where  an energy gap
pops up ($\sim 0.2$\,$\mu$eV). Similarly, the Rashba
SOC results in the anticrossing between $|S^{(-)}\rangle$
and $|T_+^{(-)}\rangle$ ($B_{\rm S^{(-)}T_+^{(-)}}\sim 2.07$\,T). 
The intersecting point between  $|S^{(-)}\rangle$
and $|T_0^{(-)}\rangle$ ($B_{\rm S^{(-)}T_0^{(-)}}\sim 1.2$\,T) is simply a crossing point.

\begin{figure}[bth]
\centering
\begin{center}
\includegraphics[width=7.5cm]{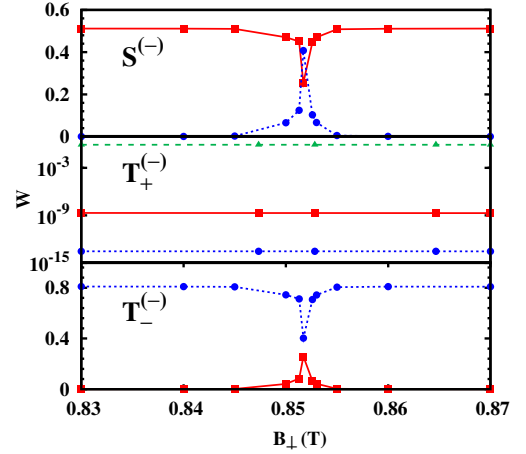}
\end{center}
\caption{(Color online) Weights of the major components in 
eigenstates, $W$ [$=|\langle
    \xi_0|\xi\rangle|^2$ with $|\xi\rangle=|S^{(-)}\rangle$, $|T_+^{(-)}\rangle$,
    $|T_-^{(-)}\rangle$, and $|\xi_0\rangle=|S^{1(-)}\rangle$ (red solid curve), 
$|T_+^{1(-)}\rangle$ (green dashed curve), $|T_-^{1(-)}\rangle$ (blue dotted curve)],
    {\em vs}. the magnetic field near the anticrossing point $B_{\rm S^{(-)}T_-^{(-)}}$.}
\label{fig3}
\end{figure}

The ST relaxation rates together with the transition rates between two
triplet states are plotted in Fig.\,\ref{fig2}(b), which
shows that the lifetimes of the excited states are extremely long
(about four orders of magnitude longer than the ST relaxation time in GaAs
QD\cite{shen}) and strongly depend on the strength of the
magnetic field. In the vicinities of 
the crossing and anticrossing points, the transition rates show
intriguing features. For example, at the anticrossing point $B_{\rm S^{(-)}T_-^{(-)}}$,
one finds that all the transition
rates except the one between $|T_+^{(-)}\rangle$ and
$|T_0^{(-)}\rangle$ present either a peak or a valley. According to the previous
works,\cite{climente,shen} the sharp decrease of
the transition rate between $|S^{(-)}\rangle$ and $|T_-^{(-)}\rangle$ results from
the decrease of the emission phonon energy. The origin of the
features of other channels can be understood from Fig.\,\ref{fig3},
which illustrates 
the major components of the states around $B_{\rm S^{(-)}T_-^{(-)}}$, e.g., $|S^{1(-)}\rangle$
(red solid curve), $|T^{1(-)}_-\rangle$ (blue dotted one) and
$|T^{1(-)}_+\rangle$ (green dashed one). 
One notices that when the magnetic field approaches $B_{\rm S^{(-)}T_-^{(-)}}$, the
composition of the $|T_+^{(-)}\rangle$ as well as
$|T_0^{(-)}\rangle$ (not shown) is almost invariant, however,
the weight of $|S^{1(-)}\rangle$ ($|T_-^{1(-)}\rangle$) in $|S^{(-)}\rangle$ 
significantly decreases (increases) due
to the spin-mixing  from the SOC.
As the component of $|T^{1(-)}_-\rangle$ in $|T_+^{(-)}\rangle$ is negligibly
small, the weight of $|S^{1(-)}\rangle$ dominates the relaxation rate.
Therefore, the relaxation rate between $|T_+^{(-)}\rangle$ and $|S^{(-)}\rangle$
decreases as shown in the inset of Fig.\,\ref{fig2}(b). However, 
the composition of
$|T_-^{(-)}\rangle$ varies in the opposite way, hence the transition rate
between $|T_+^{(-)}\rangle$ and $|T_-^{(-)}\rangle$ presents a maximum at $B_{\rm S^{(-)}T_-^{(-)}}$.
The similar feature of the channel involving $|T_0^{(-)}\rangle$ can be
interpreted in the same way. Near the anticrossing point
$B_{\rm S^{(-)}T_+^{(-)}}$, the physics is quite similar and the transition rates of all the
channels except one from $|T_0^{(-)}\rangle$ to $|T_-^{(-)}\rangle$ present either
a peak or a valley. However, in the vicinity of the crossing point
$B_{\rm S^{(-)}T_0^{(-)}}$, only the transition rate between $|S^{(-)}\rangle$ and
$|T_0^{(-)}\rangle$ shows a sharp decrease due to small phonon energy and
other transition rates change 
slightly, because there is no coupling between $|S^{(-)}\rangle$ and
$|T_0^{(-)}\rangle$ and the components of all the states remain almost
unchanged. The variation of the transition rates far way from the
  intersecting points can be understood from the dependence of the
  transition rate on the phonon energy as mentioned in the previous
  subsection.\cite{climente,shen}

\begin{figure}[bth]
\centering
\includegraphics[width=7.5cm]{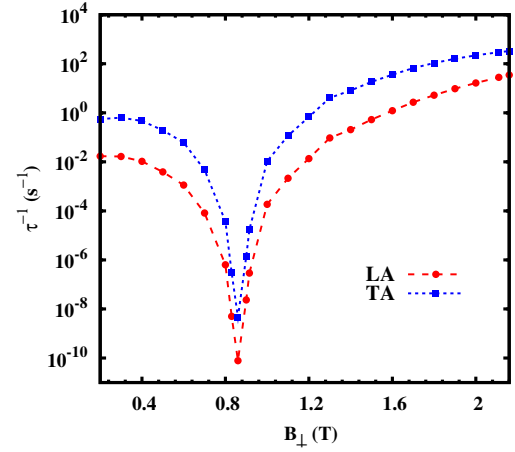}
\caption{(Color online) Contributions of the LA and TA
  phonon modes to the relaxation rate between $|T_-^{(-)}\rangle$ and
  $|S^{(-)}\rangle$. The red dashed and blue dotted curves are for the
  LA and TA modes, respectively. Here, $2a=4.344$\,nm and $d_0=29$\,nm.}
\label{fig4}
\end{figure}

To indicate the relative contribution of the LA phonon mode
to the ST relaxation, we remove the TA mode from the calculation, vice
versa. The magnetic-field dependence of the relaxation rate of the
channel between $|T_-^{(-)}\rangle$ and $|S^{(-)}\rangle$ 
is plotted in Fig.\,\ref{fig4}. One notices that the
relaxation rate of the TA mode is always larger than that of the LA
mode. Actually, the calculation of the other channels (not shown) also reveals
similar conclusion. The reason lies in the different sound velocities of the
LA and TA phonons. Since the longitudinal sound velocity  is about
twice as large as the transverse one,\cite{pop} the momentum of the LA
phonon emission  is smaller for a fixed phonon energy. As a result, the
transition rate due to the LA phonon emission process is smaller according
to Eq.\,(\ref{eq13}).

\begin{figure}
\centering
\includegraphics[width=7.5cm]{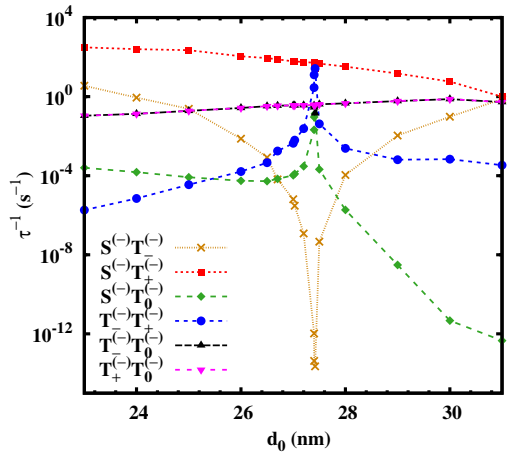}
\caption{ (Color online) Transition rates {\em vs}.
 effective dot diameter $d_0$. In the calculation, $2a=4.344$\,nm and $B_{\perp}=1$\,T.}
\label{fig5}
\end{figure}

In addition, we investigate the influence of the effective diameter
$d_0$ on the ST relaxation. The results are plotted in
Fig.\,\ref{fig5}. One notices that the behavior of
the transition rates is similar to what  obtained above
by changing the perpendicular magnetic field.  Here, an 
anticrossing point between the singlet and one of the triplets
($|T_-^{(-)}\rangle$) is also observed at $d_0\sim 27.4$\,nm. In the
vicinity of this point, we also find the relaxation rate
 between $|T_-^{(-)}\rangle$ and
$|S^{(-)}\rangle$ is strongly suppressed and the rates of other transition
channels relevant to these two states show a rapid increase or
decrease too. Therefore, the manipulation of the ST relaxation
by tuning the dot size is also feasible. In the experiment, the dot size
can be controlled electrically.\cite{shaji}

Finally, we also study the case of small valley splitting  by taking 27
monoatomic layers along the growth direction of the quantum well, where
$2|\Delta_{n_z}^1|=0.35$\,meV according to Eq.\,({\ref{eq6}}). In
this configuration, the SOC strengths are unavailable in the 
literature. We extract 
these parameters according to the results of odd monoatomic layers
calculated by Nestoklon {\em et al.}\cite{nestoklon} 
and obtain $a_0=\mp2.28$\,m/s and
$b_0=\pm37.93$\,m/s for the SOC elements between the states with identical
valley indices ``$\mp$'', when the same electric field ($30$\,kV/cm) as
the case of large valley splitting is 
applied. One finds that the lowest triplet
states (denoted as $|T_-^{(m)}\rangle$, $|T_0^{(m)}\rangle$, and
$|T_+^{(m)}\rangle$) are mainly constructed by the 
single-electron functions $|001-\rangle$ and 
$|001+\rangle$, in a QD with the effective diameter
18\,nm under a low magnetic field. 
 However, the major component of the
lowest singlet ($|S^{(-)}\rangle$) remains in the same
configuration as the case of large valley splitting. Interestingly, the
second singlet level ($|S^{\ast (m)}\rangle$), whose
major component is constructed by the single-electron basis
functions $|001-\rangle$ and $|001+\rangle$, is almost degenerate with
$|T_0^{(m)}\rangle$, which reveals that the intervalley Coulomb exchange
interaction is rather
small.\cite{li} In this case, no anticrossing point is observed
between the relevant states, because of the absence of the
SOC element between different valley states when only the lowest subband
is relevant,  as
mentioned above. Moreover, we find that the 
relaxations from the three triplet states to $|S^{(-)}\rangle$ are much
slower than those in the case of large
valley splitting because these triplets and $|S^{(-)}\rangle$ are in
  different sets as mentioned above.

\section{SUMMARY}
In summary, we have investigated the ST relaxation
  in silicon QDs with magnetic fields in either the Voigt or the 
Faraday configuration.
Our results in the Voigt configuration agree
pretty well with the recent experiment in Si/SiO$_2$ QDs. We have identified
that the origin of the relaxation channel in the experiment is 
 between the lowest singlet and triplet in the set with single
valley eigenstate ``$-/+$'' (more likely the ``$-$'' one).
 Besides, we also predict the
enhancement of the ST relaxation process in the vicinity of the
anticrossing point due to the SOCs when the magnetic field further
increases, which should be checked by future experiments.
We then focus on the ST relaxation in the Faraday configuration
 in SiGe/Si/SiGe QDs and discuss the role of the valley
splittings. In the case of large
valley splitting, the lowest levels are all constructed by the 
eigenstates from the lowest valley state.
 We find that the transition rates are about four orders of magnitude
 smaller than those of  GaAs QDs due to the weak SOC in
silicon. The transition rates can be 
effectively manipulated by tuning the magnetic field
and dot size. From the magnetic-field and dot-size dependence 
of energy levels, we also
observe ST crossing/anticrossing points. In the vicinity of
the anticrossing point, there exists a small energy gap between the
singlet and one of the triplet states due to the SOC.
The transition rates of the channels relevant to these two
states show a sharp increase or decrease. 
We  show that the contribution of the TA
phonon mode is larger than that of the LA one due to the 
smaller transverse sound velocity. As for the small valley 
splitting, the eigenstates from both valley states contribute. We find the
ST relaxation rates in this case are much smaller.

\begin{acknowledgments}
This work was supported by the Natural Science Foundation of China
under Grant No.~10725417, the National Basic Research Program of
China under Grant No.~2006CB922005 and the Knowledge Innovation
Project of Chinese Academy of Sciences. 
\end{acknowledgments}

\begin{appendix}

\section{G IN COULOMB INTERACTION}
$G$ in Eq.~(12) is given by
\begin{widetext}
\begin{equation}
G(\phi_{n_1l_1n_{z1}}^{\gamma_1},\phi_{n_2l_2n_{z2}}^{\gamma_2},
\phi_{n_1^\prime l_1^\prime n_{z1}^\prime}^{\gamma_1^\prime},
\phi_{n_2^\prime l_2^\prime n_{z2}^\prime}^{\gamma_2^\prime})
=\int_0^{\infty}dk_{\|}\int_{-\infty}^{\infty}dk_zk_{\|}
P_{n_1l_1}^{n_1^\prime l_1^\prime}(k_{\|})
P^{n_2l_2}_{n_2^\prime l_2^\prime}(k_{\|})
\frac{W_{n_{z1}\gamma_1}^{n_{z1}^\prime \gamma_1^\prime}(k_z)
(W^{n_{z2}\gamma_2}_{n_{z2}^\prime \gamma_2^\prime}(k_z))^\ast}{k^2},
\end{equation}
where $P_{nl}^{n^\prime l^\prime}$ and
$W_{n_{z}\gamma}^{n_{z}^\prime \gamma^\prime}$ come from the
lateral and vertical parts of the matrix element $\langle
n,l,n_z,\gamma|e^{i{\bf k}\cdot{\bf
    r}}|n^{\prime},l^{\prime},n_z^{\prime},\gamma^{\prime}\rangle$, respectively. $P$ 
is\cite{cheng}
\begin{equation}
P_{nl}^{n^\prime l^\prime}(k_{\|})=
\sqrt{\frac{n!n^{\prime}!}{(n+|l|)!(n^{\prime}+|l^{\prime}|)!}}{\rm
  exp}(-\frac{k_{\|}^2}{4\alpha^2})
\sum_{i=0}^{n^{\prime}}\sum_{j=0}^{n}C_{n^{\prime},|l^{\prime}|}^iC_{n,|l|}^j\bar{n}!L_{\bar{n}}^{|l-l^{\prime}|}(\frac{k_{\|}^2}{4\alpha^2})
\Big[{\rm sgn}(l^{\prime}-l)\frac{k_{\|}}{2\alpha}\Big]^{|l^{\prime}-l|},
\end{equation}
with $C_{n,l}^i=\frac{(-1)^i}{i!}{n+l \choose n-i}$ and
$\bar{n}=i+j+(|l|+|l^{\prime}|-|l^{\prime}-l|)/2$. sgn $(x)$
represents the sign function and $W$ reads
$W_{n_{z}\gamma}^{n_{z}^\prime \gamma^\prime}=\langle
n_z,\gamma|{\rm exp}(ik_zz)|n_z^{\prime},\gamma^{\prime}\rangle$.
\end{widetext}
\end{appendix}


\begin{thebibliography}{0}
\bibitem{scalable} D. Loss and D. P. DiVincenzo, Phys. Rev. A {\bf
    57}, 120 (1998).

\bibitem{koppens} F. H. L. Koppens, C. Buizert, K.-J. Tielrooij,
  I. T. Vink, K. C. Nowack, T. Meunier, L. P. Kouwenhoven, and
  L. M. K. Vandersypen, Nature (London) {\bf 442}, 766 (2006).

\bibitem{petta} J. R. Petta, A. C. Johnson, J. M. Taylor, E. A. Laird,
  A. Yacoby, M. D. Lukin, C. M. Marcus, M. P. Hanson, and
  A. C. Gossard, Science {\bf 309}, 2180 (2005).

\bibitem{koppens2} F. H. L. Koppens, J. A. Folk, J. M. Elzerman, R. Hanson,
  L. H. Willems van Beveren, I. T. Vink, H. P. Tranitz,
  W. Wegscheider, L. P. Kouwenhoven, and L. M. K. Vandersypen, Science
  {\bf 309}, 1346 (2005).

\bibitem{petta2} J. R. Petta, A. C. Johnson, A. Yacoby, C. M. Marcus,
  M. P. Hanson, and A. C. Gossard, Phys. Rev. B {\bf 72}, 161301(R)
  (2005).

\bibitem{sasaki} S. Sasaki, T. Fujisawa, T. Hayashi, and Y. Hirayama,
  Phys. Rev. Lett. {\bf 95}, 056803 (2005).

\bibitem{meunier} T. Meunier, I. T. Vink, L. H. Willems van Beveren,
  K.-J. Tielrooij, R. Hanson, F. H. L. Koppens, H. P. Tranitz,
  W. Wegscheider, L. P. Kouwenhoven, and L. M. K. Vandersypen,
  Phys. Rev. Lett. {\bf 98}, 126601 (2007).
\bibitem{cheng} J. L. Cheng, M. W. Wu, and C. L\"{u}, Phys. Rev. B
  {\bf 69}, 115318 (2004).
\bibitem{shen} K. Shen and M. W. Wu, Phys. Rev. B {\bf 76}, 235313 (2007).
\bibitem{jiang}J. H. Jiang, Y. Y. Wang, and M. W. Wu, Phys. Rev. B {\bf 77}, 
035323 (2008). 
\bibitem{golovach} V. N. Golovach, A. Khaetskii, and D. Loss,
  Phys. Rev. B {\bf 77}, 045328 (2008).
\bibitem{taylor} J. M. Taylor, H.-A. Engel, W. D\"{u}r, A. Yacoby,
  C. M. Marcus, P. Zoller, and M. D. Lukin, Nature Phys. {\bf 1}, 177
  (2005).
\bibitem{taylor2} J. M. Taylor, J. R. Petta, A. C. Johnson, A. Yacoby,
  C. M. Marcus, and M. D. Lukin, Phys. Rev. B {\bf 76}, 035315 (2007).
\bibitem{hanson} R. Hanson, L. P. Kouwenhoven, J. R. Petta, S. Tarucha,
  and L. M. K. Vandersypen, Rev. Mod. Phys. {\bf 79}, 1217 (2007).

\bibitem{hanson2} R. Hanson, B. Witkamp, L. M. K. Vandersypen,
  L. H. Willems van Beveren, J. M. Elzerman, and L. P. Kouwenhoven,
  Phys. Rev. Lett. {\bf 91}, 196802 (2003).
\bibitem{amasha} S. Amasha, K. MacLean, I. Radu,
  D. M. Zumb\"{u}hl, M. A. Kastner, M. P. Hanson, and A. C.
 Gossard, arXiv:0607110.

\bibitem{climente} J. I. Climente, A. Bertoni, G. Goldoni, M. Rontani, and
  E. Molinari, Phys. Rev. B {\bf 75}, 081303(R) (2007).
\bibitem{paget} D. Paget, G. Lampel, B. Sapoval, and
  V. I. Safarov. Phys. Rev. B {\bf 15}, 5780 (1977).

\bibitem{pikus} G. E. Pikus and A. N. Titkov, {\it Optical Orientation}
  (Berlin, Springer, 1984).
\bibitem{erlingsson} S. I. Erlingsson, Y. V. Nazarov, and V.
  I. Fal'ko, Phys. Rev. B {\bf 64}, 195306 (2001).
\bibitem{khaetskii} A. Khaetskii, D. Loss, and L. Glazman, Phys. Rev. B
  {\bf 67}, 195329 (2003).
\bibitem{witzel} W. M. Witzel and S. D. Sarma, Phys. Rev. B {\bf 74},
  035322 (2006).
\bibitem{yao} W. Yao, R.-B. Liu, and L. J. Sham, Phys. Rev. B {\bf
    74}, 195301 (2006).
\bibitem{zhang} W. Zhang, V. V. Dobrovitski, K. A. Al-Hassanieh,
  E. Dagotto, and B. N. Harmon, Phys. Rev. B {\bf 74}, 205313 (2006).
\bibitem{deng} C. Deng and X. Hu, Phys. Rev. B {\bf 78}, 245301
  (2008).
\bibitem{coish1} W. A. Coish, J. Fischer, and D. Loss, Phys. Rev. B {\bf
  77}, 125329 (2008).
\bibitem{cywinski} L. Cywi\'{n}ski, W. M. Witzel, and S. D. Sarma,
  Phys. Rev. Lett. {\bf 102}, 057601 (2009).
\bibitem{cywinski2} L. Cywi\'{n}ski, W. M. Witzel, and S. D. Sarma,
  Phys. Rev. B {\bf 79}, 245314 (2009).
\bibitem{dresselhaus} G. Dresselhaus, Phys. Rev. {\bf 100}, 580
  (1955).
\bibitem{rashba} E. I. Rashba, Fiz. Tverd. Tela (Leningrad) {\bf 2}, 1224
  (1960) [Sov. Phys. Solid State {\bf 2}, 1109 (1960)].
\bibitem{culcer} D. Culcer, L. Cywi\'{n}ski, Q. Li, X. Hu, and
  S. D. Sarma, Phys. Rev. B {\bf 80}, 205302 (2009).
\bibitem{li} Q. Li, L. Cywi\'{n}ski, D. Culcer, X. Hu, and
  S. D. Sarma, Phys. Rev. B {\bf 81}, 085313 (2010).
\bibitem{prada} M. Prada. R. H. Blick, and R. Joynt, Phys. Rev. B {\bf
    77}, 115438 (2008).
\bibitem{pan} W. Pan, X. Z. Yu, and W. Z. Shen, Appl. Phys. Lett. {\bf
  95}, 013103 (2009).
\bibitem{liu} H. W. Liu, T. Fujisawa, Y. Ono, H. Inokawa, A. Fujiwara,
  K. Takashina, and Y. Hirayama, Phys. Rev. B {\bf 77}, 073310 (2008).
\bibitem{shaji} N. Shaji, C. B. Simmons, M. Thalakulam, L. J. Klein,
  H. Qin, H. Luo, D. E. Savage, M. G. Lagally, A. J. Rimberg,
  R. Joynt, M. Friesen, R. H. Blick, S. N. Coppersmith, and
  M. A. Eriksson, Nature Phys. {\bf 4},  540 (2008).
\bibitem{culcer2} D. Culcer, L. Cywi\'{n}ski, Q. Li, X. Hu, and
  S. D. Sarma, arXiv:1001.5040.
\bibitem{xiao} M. Xiao, M. G. House, and H. W. Jiang,
  Phys. Rev. Lett. {\bf 104}, 096801 (2010).
\bibitem{coish2} W. A. Coish and D. Loss, Phys. Rev. B {\bf 70},
  195340 (2004).
\bibitem{taylor3} J. M. Taylor, W. D\"{u}r, P. Zoller, A. Yacoby,
  C. M. Marcus, and M. D. Lukin, Phys. Rev. Lett. {\bf 94}, 236803
  (2005).
\bibitem{vervoort} L. Vervoort, R. Ferreira, and P. Voisin, Phys. Rev. B
  {\bf 56}, R12744 (1997).
\bibitem{vervoort2} L. Vervoort, R. Ferreira, and P. Voisin,
  Semicond. Sci. Technol. {\bf 14}, 227 (1999).
\bibitem{nestoklon} M. O. Nestoklon, E. L. Ivchenko, J.-M. Jancu, and
  P. Voisin, Phys. Rev. B {\bf 77}, 155328 (2008).
\bibitem{boykin} T. B. Boykin, G. Klimeck, M. A. Eriksson,
  M. Friesen, S. N. Coppersmith, P. von Allmen, F. Oyafuso, and
  S. Lee, Appl. Phys. Lett. {\bf 84}, 115 (2004).
\bibitem{friesen} M. Friesen, S. Chutia, C. Tahan, and
  S. N. Coppersmith, Phys. Rev. B {\bf 75}, 115318 (2007).

\bibitem{fock} V. Fock, Z. Phys. {\bf 47}, 446 (1928).
\bibitem{darwin} C. G. Darwin, Proc. Cambrige Philos. Soc. {\bf 27}, 86
  (1931).
\bibitem{saraiva} A. L. Saraiva, M. J. Calder\'{o}n, X. Hu,
  S. D. Sarma, and B. Koiller, Phys. Rev. B {\bf 80}, 081305(R)
  (2009).
\bibitem{chutia} S. Chutia, S. N. Coppersmith, and M. Friesen,
  Phys. Rev. B {\bf 77}, 193311 (2008).

\bibitem{pop} E. Pop, R. W. Dutton, and K. E. Goodson,
  J. Appl. Phys. {\bf 96}, 4998 (2004).
\bibitem{sonder} E. Sonder and D. K. Stevens, Phys. Rev. {\bf 110},
  1027 (1958).
\bibitem{dexter} R. N. Dexter, B. Lax, A. F. Kip, and G. Dresselhaus,
  Phys. Rev. {\bf 96}, 222 (1954).

\bibitem{graeff} C. F. O. Graeff, M. S. Brandt, M. Stutzmann,
  M. Holzmann, G. Abstreiter, and F. Sch\"{a}ffler, Phys. Rev. B {\bf
    59}, 13242 (1999).
\bibitem{sze} S. M. Sze, {\it Physics of Semiconductor Devices}
  (Wiley-Interscience, New York, 1981), p. 849.


\end{thebibliography}
\end{document}